\documentclass[
 reprint,
 amsmath,amssymb,
 aps,
 prl,
]{revtex4-1}
\usepackage[T1]{fontenc}
\usepackage[utf8]{inputenc}
\usepackage{lmodern}
\usepackage[english]{babel}
\usepackage[autostyle]{csquotes}
\usepackage{refcount}
\usepackage{graphicx}
\usepackage{amsmath}
\usepackage{subfig}
\usepackage{xcolor}
\newcommand{\edit}[1]{\textcolor{black}{#1}}
\begin{document}

%\title{Structural and electronic properties of (BEDT-TTF)$_2$X$_3$, organic charge transfer salts in the $\alpha$-, $\beta$- phase X$\rightarrow$I, and $\kappa$-phases X$\rightarrow$(I,F,Br,Cl)} 
\title{Structural and electronic properties of $\alpha,\beta$-(BEDT-TTF)$_2$I$_3$ and $\kappa$-(BEDT-TTF)$_2$X$_3$ (X=I,F,Br,Cl) organic charge transfer salts} 
\author{Benjamin Commeau$^1$}
\email{benjamin.commeau@uconn.edu}
\author{R. Matthias Geilhufe$^2$}
\author{Gayanath W. Fernando$^1$}
\author{Alexander V. Balatsky$^{2,3}$}
\affiliation{$^1$University of Connecticut, Storrs, Connecticut, USA\\
$^2$Nordita, KTH Royal Institute of Technology and Stockholm University, Roslagstullsbacken 23, SE-106 91 Stockholm, Sweden\\
$^3$Institute for Materials Science, Los Alamos National Laboratory, Los Alamos, NM 87545, USA}

\date{\today}

\begin{abstract}
 (BEDT-TFF)$_2$I$_3$ charge transfer salts are reported to show superconductivity and pressure induced quasi two-dimensional Dirac cones at the Fermi level. By performing state of the art ab initio calculations in the framework of density functional theory, we investigate the structural and electronic properties of the three structural phases $\alpha$, $\beta$ and $\kappa$. \edit{We furthermore report about the irreducible representations of the corresponding electronic band structures, symmetry of their crystal structure, and discuss the origin of band crossings. Additionally, we discuss the chemically induced strain in $\kappa$-(BEDT-TTF)$_2$I$_3$ achieved by replacing the Iodine layer with other Halogens: Fluorine, Bromine and Chlorine. In the case of $\kappa$-(BEDT-TTF)$_2$F$_3$, we identify topologically protected crossings within the band structure. These crossings are forced to occur due to the non-symmorphic nature of the crystal.} The calculated electronic structures presented here are added to the organic materials database (OMDB).
\end{abstract}

\maketitle
\section{Introduction}\label{introduction}
Organic charge transfer salts, and especially, (BEDT-TTF)$_2$I$_3$ and its derivatives, have attracted the research community for almost half a century \cite{CS9912000355,liu2016insulating,nakamura2017}. Depending on the structural phase, the main reasons lie in the existence of strong many-body correlation effects due to flat bands and the opportunity of tuning the band structure by applying pressure due to the softness of the material. To date, (BEDT-TTF)$_2$I$_3$ was synthesized in five different structural phases, the $\alpha$-, $\beta$-, $\kappa$-, $\theta$- and $\lambda$-phases \cite{seo2004toward,endres1986x,schultz1986,Kobayashi87}, where the $\alpha$-, $\beta$-, and $\kappa$-phases attracted the majority of the attention. Other related compounds such as $\kappa$-(BEDT-TTF)$_2$Cu$_2$CN$_3$ have been discussed with regard to magnetic frustration in quantum spin liquids \cite{yshimizu2003}.

The $\alpha$-phase represents a narrow band-gap semiconductor at ambient pressure \cite{tajima2006electronic} showing charge ordering as investigated by synchrotron X-ray diffraction measurements \cite{kakiuchi2007charge}. Under high pressure, it was reported that $\alpha$-(BEDT-TTF)$_2$I$_3$ exhibits a transition to a semi-metallic phase having tilted Dirac-crossings within the band structure \cite{liu2016insulating,hirata2016observation,kajita2014molecular,morinari2014possible,miyahara2014possible,kondo2009crystal,mori2009requirements}. The physics of Dirac materials, i.e., materials where elementary excitations behave as effective Dirac particles \cite{wehling2014dirac}, has been of major interest during the past decades. However, besides theoretical predictions \cite{geilhufe2017three,geilhufe2016data}, $\alpha$-(BEDT-TTF)$_2$I$_3$ under high pressure is one of the few non-planar organic Dirac materials known to date. Furthermore, for the high pressure phase, superconductivity was observed with a transition temperature of 7 K \cite{Tajima2002SC}. Besides applying pressure, metallic conductivity can be achieved by illuminating the crystal with light \cite{tajima2005photo}.   

In contrast to the $\alpha$-phase, $\beta$-(BEDT-TTF)$_2$I$_3$ and $\kappa$-(BEDT-TTF)$_2$I$_3$ are metallic. Superconductivity of the $\beta$-phase was reported to depend drastically on the applied pressure having transition temperatures up to about 8 K \cite{kamarchuk1990intramolecular,Baram1986,Ginodman1985,murata1985superconductivity,creuzet1985homogeneous}. The superconducting transition temperature of the $\kappa$-phase at ambient pressure is approximately 3.4 K \cite{kobayashi1987crystal,kato1987new}. Molecular modifications of the material can increase the transition temperature to 10.4 K \cite{urayama1988new}.

Electronic structure calculations of the different phases based on tight-binding have been reported before \cite{kobayashi1987crystal,mori2009requirements}. The influence of uniaxial strain and the charge ordering in the $\alpha$-phase were \edit{also widely investigated using ab-initio methods, e.g. by Kino and Miyazaki \cite{kino2006first} or Alemany \textit{et al.} \cite{alemany2012essential}}. Correlation effects of the electrons, e.g., to describe the charge ordering in the $\alpha$-phase were discussed in terms of the Hubbard model \cite{Tanaka2016}. \edit{In the first part of the paper, we present and review electronic structure calculations for $\alpha$-, $\beta$- and $\kappa$-(BEDT-TTF)$_2$I$_3$ obtained by ab initio calculations based on density functional theory \cite{Hohenberg1964,Kohn1965,Jones2015}. Our} results complement band structure calculations performed so far by adding detailed information of the symmetry of the states in terms of the irreducible representations. This information is important for the discussion of the gap closing in the high-pressure phase of  $\alpha$-(BEDT-TTF)$_2$I$_3$ and for the investigation of line-nodes occurring on the Brillouin zone boundary of $\kappa$-(BEDT-TTF)$_2$I$_3$. \edit{In the second part, we investigate the  influence of applying chemically induced strain to the $\kappa$-phase of (BEDT-TTF)$_2$I$_3$. This is achieved by means of replacing Iodine atoms by Fluorine, Chlorine and Bromine. It turns out that for Fluorine, significant changes can be observed that are conducive for the formation of Dirac nodes slightly above the Fermi level. These crossings are formed by bands belonging to different irreducible representations, usually referred to as accidental crossings \cite{herring1937}. However, as we show in the present case, the occurrence of these crossings is not accidental, but connected to the ordering of states at the $\Gamma$-point and mediated by the connectivity of bands within the Brillouin-zone. This finding contributes to the ongoing discussion about topological band theory \cite{bradlyn2017topological}. Furthermore, all calculated electronic structures presented here are added to the organic materials database (OMDB) \cite{borysov2017organic}, which is freely accessible at \url{http://omdb.diracmaterials.org}.}

\section{Computational details}\label{computation_details}
Calculations were performed in the framework of the density functional theory~\cite{Hohenberg1964,Kohn1965,Jones2015} by applying a pseudopotential projector augmented-wave method~\cite{hamann1979norm,blochl1994projector,pseudo1,pseudo2}, as implemented in the Vienna Ab initio Simulation Package (VASP)~\cite{vasp1,vasp2,vasp3,kresse1999ultrasoft} and the Quantum ESPRESSO code~\cite{qespresso}. The exchange-correlation functional was approximated by the generalized gradient approximation according to Perdew, Burke and Ernzerhof~\cite{perdew1996} (PBE). The structural information was taken from the Cambridge Structural Database (CSD) \cite{Groom:bm5086,endres1986x,schultz1986,Kobayashi87} and transferred into POSCAR files using VESTA (Visualization for Electronic and STructural Analysis) \cite{momma2011vesta}. 

Within VASP, the precision flag was set to ``normal'', meaning that the energy cut-off is given by the maximum of the specified maxima for the cut-off energies within the POTCAR files (for example, for carbon this value is given by 400~eV). The calculations were performed spin-polarized but without spin-orbit coupling, which is a reasonable approximation due to the light elements within the (BEDT-TTF)$_2$I$_3$ molecule. For the integration in $\vec{k}$-space, a $6\times6\times6$ $\Gamma$-centered mesh according to Monkhorst and Pack~\cite{monkhorst1976special} was chosen during the self-consistent cycle. A structural optimization was performed using VASP by allowing the ionic positions, the cell shape, and the cell volume to change ($\textsc{isif}=3$). The structural optimization was performed in two ways: first, using the PBE exchange correlation functional and, second, by additionally incorporating Van der Waals corrections to the total energy according to \cite{klimes2011,klimes2010,lee2010,perez2009,Dion2004} (PBE+VdW). Quantum ESPRESSO was applied to estimate the associated irreducible representations of the energy levels within the band structure. The cut-off energy for the wave function was chosen to be 48~Ry and the cut-off energy for the charge density and the potentials was chosen to be 316~Ry. The calculated band structures using VASP and Quantum ESPRESSO are in perfect agreement. 

\section{Crystal structure and symmetry}\label{crystal structure and symmetry}
\begin{figure}[ht!]
\subfloat[Molecular ordering for the $\alpha$-, $\beta$- and $\kappa$-phase]{
\includegraphics[width=8cm]{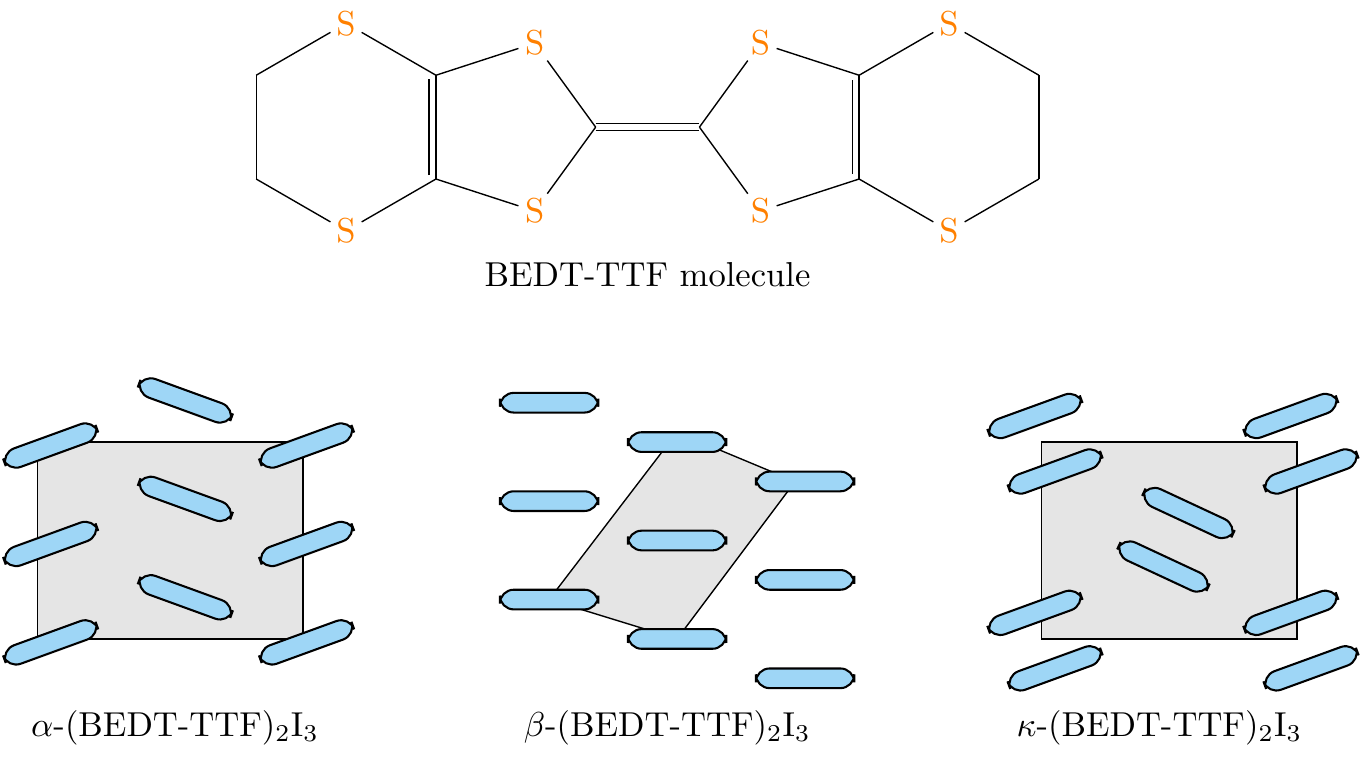}}\\
\subfloat[Layered structure of $\alpha$-(BEDT-TTF)$_2$I$_3$ \cite{momma2011vesta}\label{crystal_layer}]{
\includegraphics[width=5.5cm]{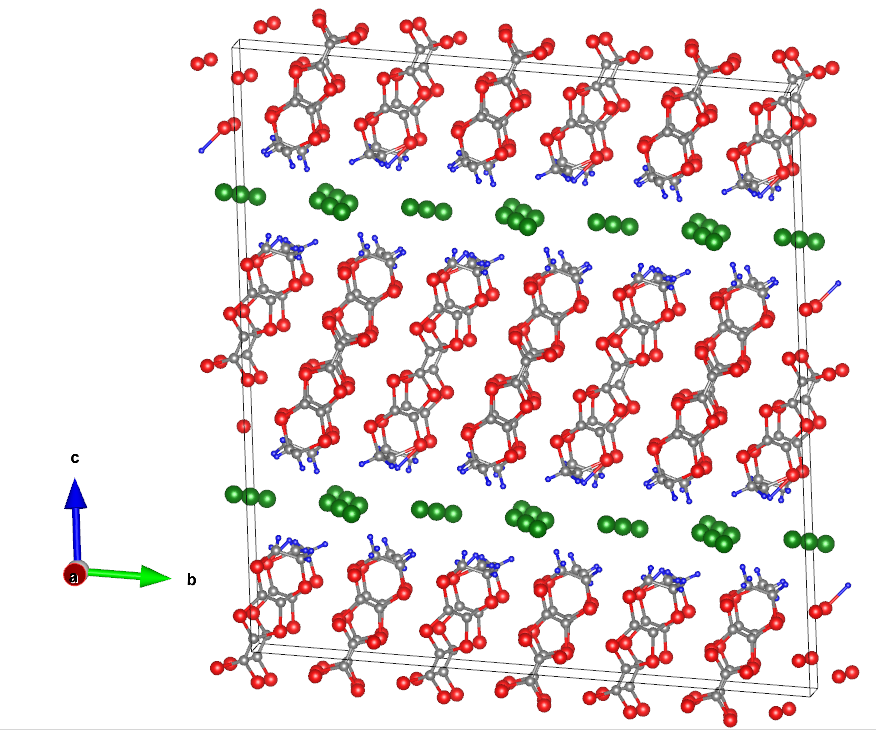}}
\caption{Crystal structure for different structural phases of (BEDT-TTF)$_2$I$_3$}
\label{crystal_structure}
\end{figure}
\begin{table}[t!]
\caption{Experimental and computational lattice constants of $\alpha$-, $\beta$- and $\kappa$-(BEDT-TTF)$_2$I$_3$. The experimental lattice constants were used as a starting configuration for the structural optimization.}
\begin{tabular}{lcccccc}
\hline\hline
 & $a$ & $b$ & $c$ & $\alpha$ & $\beta$ & $\gamma$ \\
 \hline
 $\alpha$-phase (PBE) & 9.16&10.77 &17.54 & 96.73&97.58 & 91.47\\
 $\beta$-phase (PBE) & 6.91 & 9.79 & 15.82 & 90.7 & 93.1 & 109.2 \\
  $\kappa$-phase (PBE) & 16.86 &  9.39 & 13.53 & 90.0 & 105.2 & 90.0 \\
  \hline
  $\alpha$-phase (PBE+VdW) & 9.15  & 10.75 & 17.53 & 96.50 & 97.76 & 90.64 \\
 $\beta$-phase (PBE+VdW) & 6.56 & 9.12 & 15.33 & 94.96 & 96.93 & 110.61 \\
  $\kappa$-phase (PBE+VdW) & 16.50 & 8.48 & 12.89 & 90.00 & 108.86 & 90.00 \\
  \hline
 $\alpha$-phase (exp.) \cite{endres1986x}  & 9.07 & 10.72 & 17.40 & 96.6 & 97.8 & 91.1 \\
 $\beta$-phase (exp.) \cite{schultz1986} & 6.59 & 9.04 & 15.20 & 94.9 & 95.7 & 110.0 \\
 $\kappa$-phase (exp.) \cite{Kobayashi87} & 16.43 & 8.68 & 13.06 & 90.0 & 107.9 & 90.0 \\
 \hline 
 \hline
\end{tabular}
\label{lattice_parameters}
\end{table}
 \begin{figure*}[t!]
 \subfloat[$\alpha$-phase band structure and DOS\label{band_structure_alpha}]{
 \includegraphics[width=8.5cm]{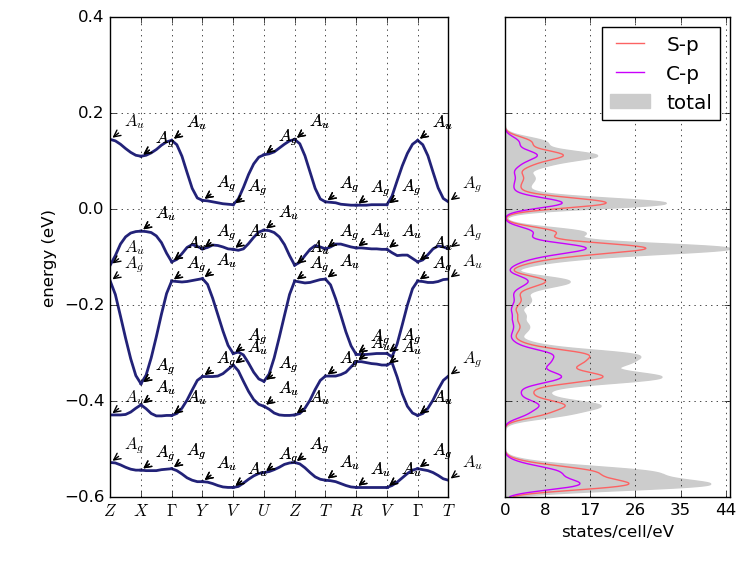}}\hfill
 \subfloat[$\beta$-phase band structure and DOS\label{band_structure_beta}]{
 \includegraphics[width=8.5cm]{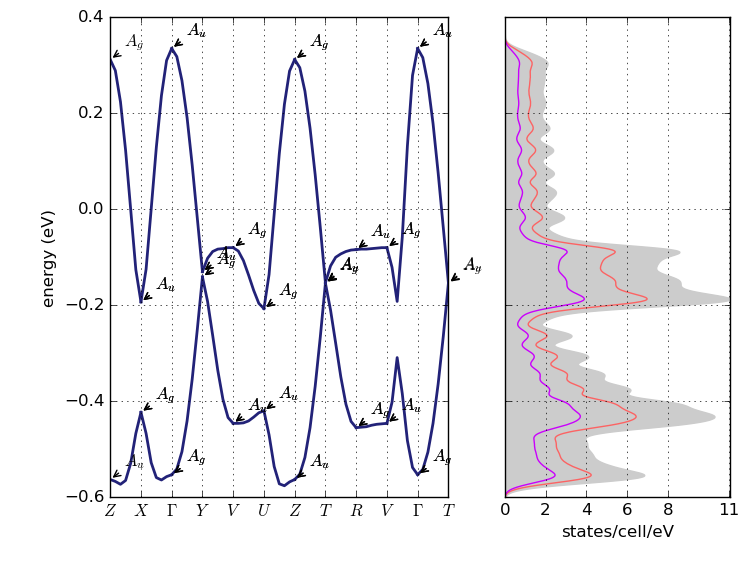}}\\
 \subfloat[$\kappa$-phase band structure and DOS\label{band_structure_kappa}]{
 \includegraphics[width=8.5cm]{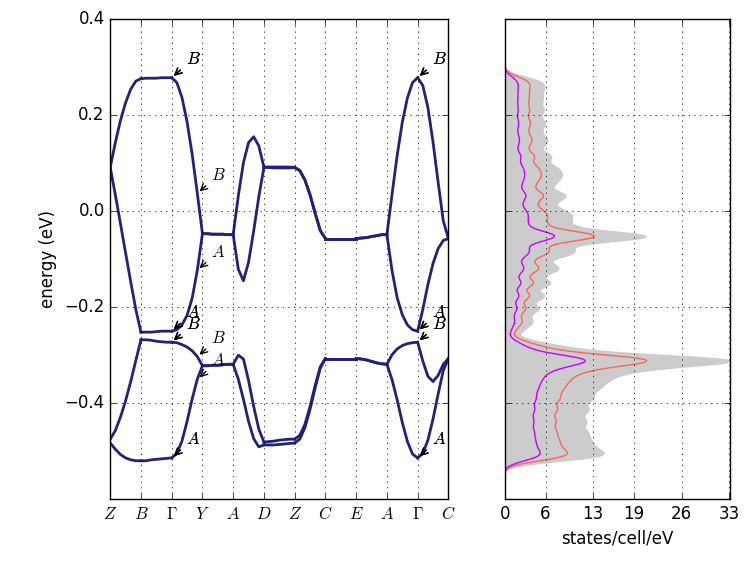}}\hfill
 \subfloat[\edit{triclinic Brillouin zone}\label{bz_tc}]{\raisebox{1cm}{
 \includegraphics[width=4cm]{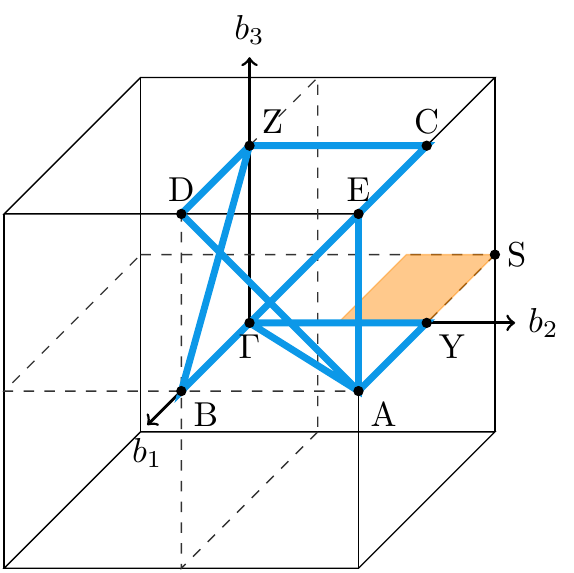}}}\hspace{0.2cm}
 \subfloat[monoclinic Brillouin zone\label{bz_mc}]{\raisebox{1cm}{
 \includegraphics[width=4cm]{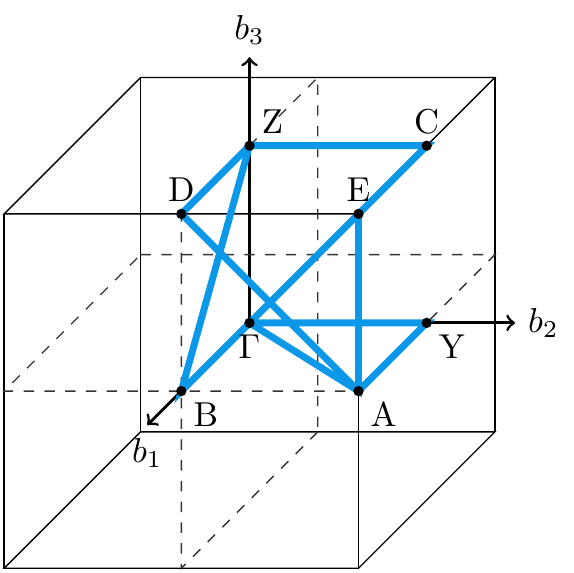}}}\\
 \caption{Band structure, irreducible representations and density of states of $\alpha$-, $\beta$- and $\kappa$-(BEDT-TTF)$_2$I$_3$. The Brillouin zones are plotted in units of the reciprocal lattice vectors and show the paths for the band structure calculations. \edit{The orange plane in (d) indicates the region for the band structure plot in Figure \ref{alpha2D}.}\label{band_structure}}
 \end{figure*}
Pictures of the crystal structures and the molecular ordering within the unit cells of $\alpha$-, $\beta$- and $\kappa$-(BEDT-TTF)$_2$I$_3$ are provided in Fig. \ref{crystal_structure}. The $\alpha$- and the $\beta$-phase of (BEDT-TTF)$_2$I$_3$ crystallize in a triclinic crystal structure, having the space group $P\overline{1}$ ($\#$2). Since for every space group $\mathcal{G}$, the group of primitive lattice translations $\mathcal{T}$ forms an invariant subgroup, $\mathcal{G}$ can be represented in terms of a coset decomposition with respect to $\mathcal{T}$. In the case of $P\overline{1}$ only two coset representatives are present, which are the identity $E$ and the inversion $I$,
\begin{equation}
P\overline{1} = E \mathcal{T} + I \mathcal{T}.
\label{eq1}
\end{equation}

i.e., the corresponding space group, ($\mathcal{G},\odot$), expressed as a union of the cosets, is $$\mathcal{G} = \mathcal{T}\cup \left\{I|(0,0,0)\right\}\odot\mathcal{T}.$$
The $\kappa$-phase crystallizes in a monoclinic crystal structure having the space group $P12_1 1$ ($\#$4), which can be represented by
\begin{equation}
P12_1 1 = E \mathcal{T} + \left\{C_{2y}|(0,1/2,0)\right\} \mathcal{T}.
\label{eq2}
\end{equation}
Here, $\left\{C_{2y}|(0,1/2,0)\right\}$ denotes a screw, represented by a two-fold rotation about the $y$-axis together with a non-primitive shift along the lattice vector $\vec{a}_2$. The translation group $\mathcal{T}$ in equations \eqref{eq1} and \eqref{eq2} is not similar since it depends on the individual lattice. 

Even though $\alpha$-, $\beta$- and $\kappa$-(BEDT-TTF)$_2$I$_3$ are three-dimensional crystals, the crystal structures exhibit a stacking of (BEDT-TTF)$_2$I$_3$-layers separated by iodine atoms (see Fig. \ref{crystal_layer} for $\alpha$-phase). Whereas the electron hopping is reasonably strong between molecules within each layer, the interlayer coupling is weak, leading to quasi-2d band structure \cite{katayama2006pressure}. The individual lattice parameters for the three phases before and after the structural optimization using VASP are shown in Table \ref{lattice_parameters}.

Starting from the experimental lattice parameters and atomic coordinates, each phase was allowed to relax using forces calculated from first principles. The $\kappa$-phase was relaxed until the magnitude of force components on each atom was less than 0.05 eV/$\AA$.
The final calculated pressure was of the order of 0.2 kbar. The relaxations carried out using VASP (ISIF=3) did not alter the crystalline symmetry. In comparison to the experimental values, the unit cell volume was increased for both, the standard PBE run and PBE+VdW. 

\edit{\section{Electronic structure of (BEDT-TTF)$_2$I$_3$\label{equilibrium electronic structure}}}
The results of the electronic structure calculations for the three investigated structural phases of (BEDT-TTF)$_2$I$_3$ are shown in Fig. \ref{band_structure}. As can be verified in the density of states for all phases, the main contribution to the electronic states close to the Fermi level comes from the $p$-electrons of sulfur and carbon. In general, the DFT calculations reveal the charge transfer nature of the compound where sulfur and hydrogen donate charge whereas carbon and iodine accept charge, giving charge differences in the range of 0.2-0.4 electrons per site. 

\begin{figure}[h!]
\includegraphics[width=6cm]{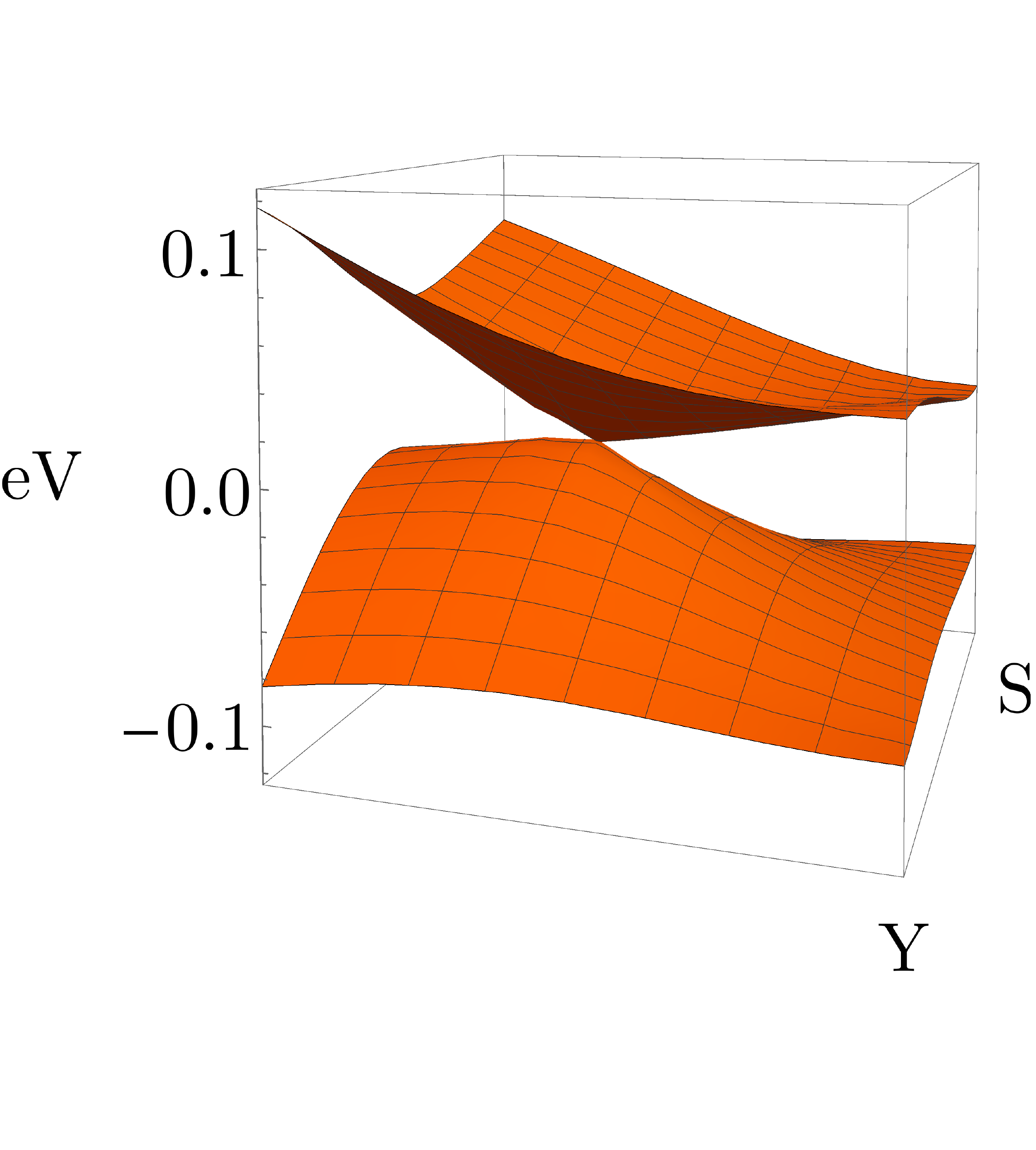}
\caption{\edit{Tilted linear crossing within the band structure of $\alpha$-(BEDT-TTF)$_2$I$_3$ under a pressure of 1.76 GPa. The plot region within the Brillouin zone is shown in Figure \ref{bz_tc}.}\label{alpha2D}}
\end{figure}

The equilibrium band structure of the $\alpha$-phase shows an indirect narrow band gap in the order of 0.05 eV. Except at the $\Gamma$-point, the little co-group of each $\vec{k}$-point within the interior of the Brillouin zone  \cite{bradley2010mathematical} is given by the trivial group, i.e., the group containing the identity only \cite{aroyo2006bilbao,HergertGeilhufe}. Hence, states belonging to each band within the interior transform as the identity representation. A distinction between even ($A_g$) and odd ($A_u$) states with respect to the inversion can only be made on the $\Gamma$-point and other high symmetry points on the Brillouin zone boundary.
\edit{Under high pressure, $\alpha$-(BEDT-TTF)$_2$I$_3$ establishes a tilted linear crossing of bands as can be seen in Fig. \ref{alpha2D}. The picture illustrates the band structure of $\alpha$-(BEDT-TTF)$_2$I$_3$ under a pressure of 1.76 GPa. The related lattice constants were chosen according to \cite{kondo2009crystal}. This crossing occurs even though both touching bands transform as the same irreducible representation and hybridization should be expected. The (tilted) Dirac Point in Fig. \ref{alpha2D} is located at $\vec{k}_1=(-0.225, 0.35)$. Such band crossings (or contacts) can happen in two-dimensional systems at a general (incommensurate) $\vec{k}$-point with the contact satisfying a feasibility ({\sl or generalized von Neumann-Wigner}) condition, as discussed in Ref.~\cite{asano-hotta}. As a consequence of the crystalline symmetry, a second Dirac crossing can be observed at the inverted $\vec{k}$-point $\vec{k}_2=(0.225, -0.35)$. The high symmetry points Y and S denote $\vec{k}_Y=(0.0, 0.5,0.0)$ and $\vec{k}_S=(-0.5, 0.5,0.0)$, respectively.}

Since the $\beta$-phase has the same space group symmetry as the $\alpha$-phase, the same kinds of irreducible representations occur on the energy bands. However, as can be seen in Fig. \ref{band_structure_beta}, the band structure and density of states clearly show a metallic material. The density of states of the $\beta$-phase remains almost constant for energies above the Fermi level.

A metallic behavior is also found for the $\kappa$-phase. However, due to the nonsymmorphic space group, degeneracies of energy bands can be observed at the Brillouin zone boundary belonging to pairs of complex conjugate one-dimensional irreducible representations \cite{bradley2010mathematical}. These degenerate states can be referred to as line-nodes protected by the crystalline symmetry \cite{geilhufe2017three}. Regarding the path $\overline{\Gamma Y}$ in the Brillouin zone, wave functions can transform either as even ($A$) or odd ($B$) under the two-fold rotation. However, at the point $Y$ one even and one odd band have to merge. This leads to a general pattern within the band structure of materials having the space group $P12_1 1$ showing pairs of bands sticking together. 

\section{Chemical Strain Induced Changes}
The (BEDT-TTF) layers and the anion layer are stacked alternately in these materials and the latter is thought to be insulating. The bands near the Fermi level are mostly originating from (BEDT-TTF) layers. In the $\kappa$-phase, there is a set of bonding (occupied) and antibonding (partially occupied) bands in the vicinity of the Fermi level. We have examined the role of the anion layer as follows. The $\kappa$-phase of (BEDT-TTF)$_2$I$_3$ was studied by replacing all the Iodine atoms  with another Halogen atom, namely Bromine, Chlorine, and Fluorine. These atoms are isovalent and smaller in size compared to Iodine and hence can induce "chemical strains." Such chemical substitutions could be regarded as giving rise to a compression of the original unit cell containing Iodine. We have monitored the relaxed volumes of the unit cell under such substitutions and found a steady decrease in the volume of the unit cell under the progressive substitutions I$\rightarrow$Br$\rightarrow$Cl$\rightarrow$F (see Table \ref{lattice_parameters_X}). This volume contraction is noticeably stronger with the inclusion of VdW forces. In addition, for Bromine and Chlorine substitutions, main features of the band structure, such as the (filled) bonding and (partially filled) antibonding combinations near the Fermi level, appear qualitatively unchanged compared to Figure \ref{band_structure_kappa}. The band widths are slightly enhanced, as one might expect, due to volume contraction.
 
\begin{figure}[!h]
\includegraphics[width=8.cm]{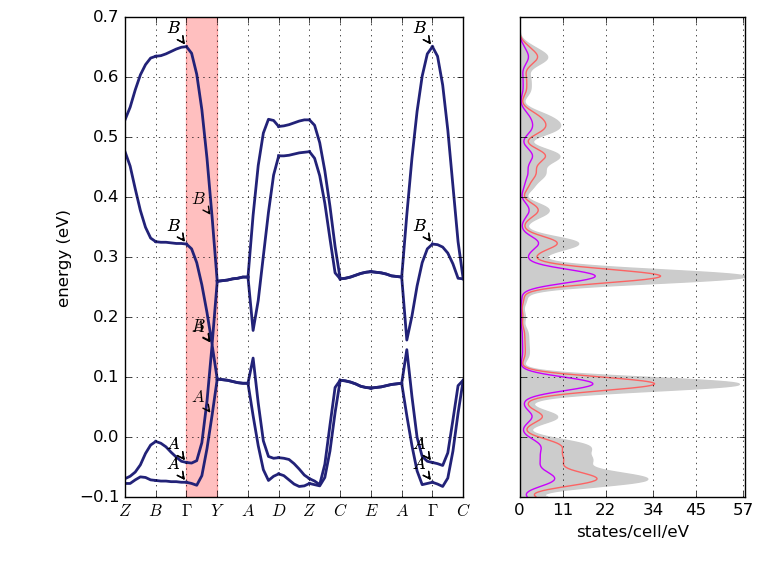}
\caption{Band structure, irreducible representations and density of states for $\kappa$-(BEDT-TTF)$_2$F$_3$. Path $\overline{\Gamma Y}$ is highlighted in pink. \label{band_structure_substitution}}
\end{figure}

However, with the $\kappa$-phase Fluorine substitution, there are significant changes in the unit cell volume and band structure around the Fermi level as can be seen in Table \ref{lattice_parameters_X} and Figure \ref{band_structure_substitution}. The unit cell volume has decreased to about 80\% of what it was with Iodine. This is not completely unexpected since Fluorine is the smallest Halogen atom. It appears that the aforementioned bonding and antibonding set of bands has been shifted up with respect to the Fermi level. During the self-consistent iterations with structural relaxations, the system appears to go through a metastable magnetic phase. However, the completely relaxed structure shows no net magnetic moment while, after full relaxation, the forces on each atom are less than 0.05 eV/$\AA$. 
These results are quite relevant and interesting in the context of magnetism, quantum spin liquids and strong correlations since some of the bands that crossed the Fermi level are quite flat, hence have peaks in the density of states. In addition, it shows that controlling and tuning the anion part of this molecule can give rise to many fascinating phenomena, as discussed in some recent publications \cite{liu2016insulating,nakamura2017}.

\begin{table*}[ht!]
\caption{Computed lattice constants of $\kappa$-(BEDT-TTF)$_2$X$_3$, with X denoting F, Cl and Br. Experimental lattice constants of $\kappa$-(BEDT-TTF)$_2$I$_3$ were used as a starting configuration for the structural optimization. Lattice parameters are given in \AA. $V_{\text{X}}/V_{\text{I}}$ denotes the ratio of the cell volume with respect to $\kappa$-(BEDT-TTF)$_2$I$_3$ according to Table \ref{lattice_parameters}.}
\begin{tabular}{lccccccc}
\hline\hline
 & $a$ & $b$ & $c$ & $\alpha$ & $\beta$ & $\gamma$ & $V_{\text{X}}/V_{\text{I}}$ (\%) \\
  \hline
  $\kappa$-(BEDT-TTF)$_2$F$_3$ (PBE) & 14.82 & 8.59 & 12.17 & 90 & 110.49 & 90.0  & 81.25 \\
 $\kappa$-(BEDT-TTF)$_2$Cl$_3$ (PBE) & 16.09 & 8.75 & 12.98 & 90.0 & 107.89 & 90.0 &  98.14 \\
  $\kappa$-(BEDT-TTF)$_2$Br$_3$ (PBE) & 16.35 & 8.72 & 13.05 & 90.0 & 108.18 & 90.0 & 99.78 \\
   \hline
 $\kappa$-(BEDT-TTF)$_2$F$_3$ (PBE+VdW) & 14.86 & 8.62 & 11.22 & 90.0 & 102.52 & 90.0 & 82.19 \\
 $\kappa$-(BEDT-TTF)$_2$Cl$_3$ (PBE+VdW) & 15.88 & 8.45 & 12.80 & 90.0 & 108.60 & 90.0 & 95.30 \\
  $\kappa$-(BEDT-TTF)$_2$Br$_3$ (PBE+VdW) & 16.19 & 8.47 & 12.89 & 90.0 & 109.35 & 90.0  & 97.66 \\
 \hline 
 \hline
\end{tabular}
\label{lattice_parameters_X}
\end{table*}

In addition to a shift of the bands, a change of the ordering of the levels at the $\Gamma$ point occurs, highlighted in pink in Figure \ref{band_structure_substitution}. Considering the 4 bands closest to the Fermi level for $\kappa$-(BEDT-TTF)$_2$I$_3$ as shown in Figure \ref{band_structure_kappa}, the irreducible representations at $\Gamma$ have the ordering $A$ ($-0.50$ eV), $B$ ($-0.27$ eV), $A$ ($-0.24$ eV), $B$ ($0.27$ eV). However, for $\kappa$-(BEDT-TTF)$_2$F$_3$ an ordering $A$ ($-0.06$ eV), $A$ ($-0.03$ eV), $B$ ($0.37$ eV), $B$ ($0.62$ eV) is observed. Since always even and odd bands are merging at the Brillouin zone boundary (e.g. at $Y$) a crossing of an $A$ and a $B$ band is forced to occur, e.g., on the path $\overline{\Gamma Y}$. Such topological crossings were discussed before in References \cite{geilhufe2017three,geilhufe2016data,PhysRevB.94.155108,bouhon2017global,bradlyn2017topological}. Following the scope of Reference \cite{geilhufe2017three} it is possible to introduce a topological invariant
\begin{equation}
n \equiv \frac{1}{2} \sum_{i=1}^2 \chi^i_\Gamma(C_{2y})\pmod{2}
\end{equation}
where the sum runs over the lower two bands and $\chi^i_\Gamma(C_{2y})$ denotes the characters of $C_{2y}$ of the associated irreducible representations of these bands at the $\Gamma$ point. The trivial case $n=0$ is present for $\kappa$-(BEDT-TTF)$_2$I$_3$, whereas the non-trivial case $n=1$ is found for $\kappa$-(BEDT-TTF)$_2$F$_3$. Hence, the band crossing corresponds to a topologically protected Dirac point. By changing the Fermi level, e.g., by doping or gating, a Dirac semimetal state could be achieved.

\section{Conclusions}
Using first principles electronic structure methods, we have studied the $\alpha$-, $\beta$-, and $\kappa$- phases of (BEDT-TTF)$_2$I$_3$ organic charge transfer salts.
We performed structural optimization on all lattices. The computational lattice parameters changed by at most 8.3\% from experiment to relaxation, with the $\beta$-phase showing the largest change in lattice parameter size.
The $\alpha$- and the $\beta$-phase of (BEDT-TTF)$_2$I$_3$ crystallize in a triclinic crystal structure having the space group $P\overline{1}$ ($\#$2). We calculated the band structures and identified the associated irreducible representations (even and odd with respect to inversion symmetry). Our $\alpha$-phase band structure revealed an indirect band gap of 0.05 eV. In comparison, the band structure for the $\beta$-phase revealed bands crossing the Fermi level suggesting it behaves as a metal. For the $\beta$-phase, we discovered a flat density of states above its Fermi level similar to a non-interacting Fermi gas. The $\kappa$-phase crystallizes in a monoclinic crystal structure having the nonsymmorphic space group $P12_1 1$ ($\#$4). Due to that, degeneracies of energy bands can be observed at the Brillouin zone boundary belonging to pairs of complex conjugate one-dimensional irreducible representations. These degenerate states can be referred to as line-nodes protected by its crystalline symmetry.
By replacing the Iodine layer with other halides (F, Cl, Br) the band structure can be tuned by means of chemical strain. In this context, we have discovered that chemical substitution of the $\kappa$-phase from Iodine to Fluorine mimics applied pressure. With Fluorine substitution, the unit cell volume decreased by about 20\% compared to Iodine. Its bonding and antibonding set of bands has been shifted up with respect to the Fermi level. In addition, we have identified a topologically protected Dirac point here. The bands that crossed the Fermi level are quite flat and hence have peaks in the density of states. This suggests an enhancement of strong electron correlations.

Organic materials based on BEDT-TTF show great potential in fine tuning specific band structures near the Fermi level due to their softness. Our research shows that applying chemical strain to the $\kappa$ phase can create strong electron correlations due to the flatness of its band structure. 
%The general scope of research on BEDT-TTF should investigate what methods can experimentalists use to change the band structure (such as applying pressure) to create strongly correlated materials and experimentally measure what new properties are created from the strong correlations. 
The calculated electronic structures of the $\alpha$-, $\beta$- and $\kappa$-phase of (BEDT-TTF)$_2$I$_3$ presented here will be added to the organic materials database (OMDB) \cite{borysov2017organic}, which is freely accessible at \url{http://omdb.diracmaterials.org}.

\section{Acknowledgement}\label{acknowledgement}
  We are grateful to K. Kanoda and G. Montambaux  for useful discussions. The work at Los Alamos is supported by the US Department of Energy, BES E3B7. Furthermore, we are grateful for support from the Swedish Research Council Grant No.~638-2013-9243, the Knut and Alice Wallenberg Foundation, and the European Research Council under the European Union’s Seventh Framework Program (FP/2207-2013)/ERC Grant Agreement No.~DM-321031. The authors acknowledge computational resources from the Swedish National Infrastructure for Computing (SNIC) at the National Supercomputer Centre at Link\"oping University and the High Performance Computing Center North (HPC2N), the High Performance Computing (HPC) cluster at the University of Connecticut, and the computing resources provided by the Center for Functional Nanomaterials, 
 which is a U.S. DOE Office of Science Facility, at Brookhaven National Laboratory under Contract No. DE-SC0012704.

\bibliography{mybib.bib}
\end{document}